\documentclass[fleqn,twoside]{article}

\usepackage{amssymb}
\usepackage{graphicx}

\usepackage{amsmath}

\begin{document}
\noindent
{\sf University of Shizuoka}

\hspace*{13cm} {\large US-04-06}
\vspace{2mm}

\begin{center}

{\Large\bf Comments on a Flavor Symmetry } \footnote{
Talk at the YITP Workshop 2004 ``Progress in Particle
Physics", June 29 -- July 2, Kyoto University, Japan.
}

\vspace{2mm}

{\bf Yoshio Koide}

{\it Department of Physics, University of Shizuoka, 
52-1 Yada, Shizuoka 422-8526, Japan\\
E-mail address: koide@u-shizuoka-ken.ac.jp}

\end{center}



\vspace{5mm}


\noindent{\large\bf 1. Introduction} \ 

We know that the masses of the charged
fermions rapidly increase as $(u,d,e) \rightarrow
(c,s,\mu) \rightarrow (t,b,\tau)$.
At first, it has been considered that such rapid increasing
of the mass spectra cannot be understood from
an idea of ``symmetry".
The horizontal degree of freedom has been called as
``generations".
In contrast to the idea of ``generations", there is
an idea of ``families" that the horizontal quantum
number states have basically the same opportunity.
After the democratic mass matrix model [1]
was proposed, the idea of ``families" became 
one of the promising viewpoints for ``flavors".
Nowadays, a popular idea to understand the observed 
quark and lepton mass spectra and mixing matrices is 
to assume a flavor symmetry which puts constraints on
the Yukawa coupling constants.

We sometimes take a phenomenological approach 
``symmetry + its breaking" for the fermion mass 
matrices.
For example, in order to understand the neutrino
masses and mixings, we put a flavor symmetry on the
neutrino mass matrix on the flavor basis where the charged
lepton mass matrix is diagonal.
However, we must remember that the left-handed neutrino
$\nu_{Li}$ is a partner of the left-handed charged
lepton $e_{Li}$ of the SU(2)$_L$ doublet.
We usually consider that the SU(2)$_L$ symmetry is
unbroken until a low energy (electroweak) scale $M_W$.
Therefore, if we consider a flavor symmetry under which
the fields $\nu_L$ and $e_L$ are transformed as
$\nu_L =U_X \nu'_L$ and $e_L =U_X e'_L$, 
the neutrino and charged lepton mass matrices 
$M_\nu$ and $M_e$ must  satisfy the relations
$$
 U_X^\dagger M_\nu U_X^* = M_\nu, \ \ \   
U_X^\dagger M_e M_e^\dagger U_X= M_e M_e^\dagger ,
\eqno(1)
$$
for the energy scale $\mu > M_W$.
Also, the up- and down-quark mass matrices $M_u$ and
$M_d$ must satisfy the relations
$$
U_X^\dagger M_u M_u^\dagger U_X= M_u M_u^\dagger , \ \ \ 
U_X^\dagger M_d M_d^\dagger U_X= M_d M_d^\dagger .
\eqno(2)
$$

In the present talk, we will point out [2] 
that if a flavor symmetry (a discrete symmetry, 
a U(1) symmetry, and so on) exists, we cannot obtain
the observed  Cabibbo-Kobayashi-Maskawa  
(CKM) quark mixing matrix $V_q$ and  
Maki-Nakagawa-Sakata (MNS) lepton mixing 
matrix $U_\ell$, even if we can obtain reasonable 
mass spectra under the symmetry. 
Such the serious constraint is derived only from the relations
(1) [and also (2)], and we will not assume any explicit 
flavor symmetry and/or any explicit mass matrix forms.
And then, we will discuss the meaning of the severe result.

\vspace{2mm}

\noindent{\large\bf 2. What happens if a flavor symmetry exists?}\ 

First, we investigate relations in the quark sectors under 
the conditions (2).
The Hermitian matrix $M_f M_f^\dagger$ ($f=u,d$) is, in general, 
diagonalized as
$$
(U_L^f)^\dagger M_f M_f^\dagger U_L^f = D_f^2 \equiv
{\rm diag}( m_{f1}^2, m_{f2}^2,m_{f3}^2) ,
\eqno(3)
$$
and the CKM matrix $V_q$ is given by $V_q \equiv (U_L^u)^\dagger U_L^d$.
{}From Eqs.~(2) and (3), we obtain the relation 
$(U_X^f)^\dagger D_f^2 U_X^f= D_f^2$, 
where
$U_X^f = (U_L^f)^\dagger U_X U_L^f$.
Therefore, the matrix $U_X^f$  must 
be a diagonal matrix with a form
$U_X^f = P_X^f \equiv {\rm diag}( e^{i \delta_1^f},
e^{i \delta_2^f}, e^{i \delta_3^f})$,
unless the masses are not degenerated.
Therefore, we obtain
$U_X = U_L^u P_X^u (U_L^u)^\dagger = U_L^d P_X^d (U_L^d)^\dagger$,
which leads to a constraint on the CKM matrix:
$P_X^u = V_q P_X^d (V_q)^\dagger$, i.e.
$$
(e^{i \delta_i^u } -e^{i \delta_j^d }) 
(V_q)_{ij} =0  \ \ (i,j=1,2,3).
\eqno(4)
$$
Only when $\delta_i^u = \delta_j^d$, we can
obtain $(V_q)_{ij} \neq 0$.
Since we do not consider a trivial case with 
$U_X \propto {\bf 1}$, we cannot consider such a case
as all elements of $V_q$ are not zero.
We can obtain only the CKM mixing between two families.

For the lepton sectors, the situation is the same.
{}From Eqs.~(1) and (3), we again obtain a severe constraint
on the MNS mixing matrix $U_\ell$:
$$
(e^{i \delta_i^e } -e^{i \delta_j^\nu }) 
(U_\ell)_{ij} =0  \ \ (i,j=1,2,3).
\eqno(5)
$$

\vspace{2mm}
\noindent{\large\bf 3. How to evade this trouble}\ 

One way  to evade the present severe conclusions (4) and (5)
is to adopt a model with no flavor symmetry.
We consider that each generation has a hierarchically different 
structure, so that
the fermion flavors are ones which should be understood  
from the concept of ``generations" rather than 
from that of ``families".
For example, in the Froggatt and Nielsen model [3], 
the flavor has a ``generation" structure, so that the model should be
regarded as a model with ``no flavor symmetry", although 
the model is based on a U(1) symmetry at a high energy scale.
The model can evade the present trouble, and it is indeed one 
of the most promising models which can reasonably understand 
the generations.

However, we know the fact (the degree of freedom
of ``rebasing") that 
we cannot physically distinguish two mass matrix sets 
$(M_u, M_d)$ and $(M'_u, M'_d)$, where $(M'_u, M'_d)$ 
is obtained from $(M_u, M_d)$ by a common flavor-basis 
rotation for the SU(2)$_L$ doublet fields. 
(The situation is the same in the lepton sector.)
Only when there is a flavor symmetry, the mass matrix forms
$(M_u, M_d)$ in a specific flavor basis have a meaning, 
because the operator of the flavor rotation does not 
commute with the flavor symmetry operator $U_X$.
Therefore,  the idea of a flavor symmetry is still attractive
to most mass-matrix-model-builders.

In order to evade the troubles (4) and (5)
within the framework of the ``families" (not
``generations"), we have to seek for
a flavor symmetry breaking mechanism with the
following conditions:  
(i) The original Lagrangian (including the
symmetry breaking mechanism) is exactly invariant
under the SU(2)$_L$.
(ii) The flavor symmetry should be completely
broken at a high energy scale $M_X$, so that we cannot 
have any flavor symmetry below $\mu=M_X$.

For example, let us consider a two Higgs doublet model,
or a $\overline{5}_L \leftrightarrow \overline{5}'_L$ model.
In such a model, the effective Yukawa coupling constants
$Y^f$ below $\mu=M_X$ are given by a linear combination 
of two Yukawa coupling constants with different textures
$Y^f_A$ and $Y^f_B$,
$$
Y^f = c_A^f Y_A^f +c_B^f Y_B^f ,
\eqno(6)
$$
so that $Y^f (Y^f)^\dagger$ do not satisfy the flavor
symmetry conditions (1) [or (2)].
However, we should note that the matrices $Y_A^f(Y^f_A)^\dagger$ 
and $Y_B^f(Y^f_B)^\dagger$ have to satisfy the SU(2)$_L$ constraints 
individually.
Regrettably, Some of currently proposed models with a phenomenological
flavor symmetry breaking seem to be unconcerned about this
SU(2)$_L$ constraints.

\vspace{3mm}

\noindent
[1] H.~Harari, H.~Haut and J.~Weyers, Phys.~Lett. {\bf 78B} (1978) 459.

\noindent
[2] Y.~Koide, hep-ph/0406286.

\noindent
[3] C.~D.~Froggatt and H.~B.~Nielsen, Nucl.~Phys. {\bf B147} (1979) 277.

\end{document}